\pdfoutput=1 
\documentclass[12pt]{article}

\usepackage{times}
\usepackage{graphicx}
\usepackage{color}

\newenvironment{Natabstract}{%
\begin{quote} \bf}
{\end{quote}}

\newcounter{lastnote}

\title{Fe$_{3}$O$_{4}$ thin films: controlling and manipulating an elusive quantum material}

\author
{X.~H.~Liu, C.~F.~Chang, A.~D.~Rata$^{\dag}$, A.~C.~Komarek, and L.~H.~Tjeng
\\
\normalsize{Max Planck Institute for Chemical Physics of Solids,}\\
\normalsize{N\"othnitzerstr. 40, 01187 Dresden, Germany}\\
\\
\normalsize{$^{\dag}$current address: Institute of Physics, Martin Luther University,}\\
\normalsize{Halle-Wittenberg, 06099 Halle, Germany}
}

\begin{document}
\baselineskip22pt

\date{07 July, 2016}
\maketitle

\begin{Natabstract}

Fe$_3$O$_4$ (magnetite) is one of the most elusive quantum materials and at the same time
one of the most studied transition metal oxide materials for thin film applications.
The theoretically expected half-metallic behavior generates high expectations
that it can be used in spintronic devices.
Yet, despite the tremendous amount of work devoted to preparing
thin films, the enigmatic first order metal-insulator
transition and the hall mark of magnetite known as the Verwey transition,
is in thin films extremely broad and occurs at substantially lower temperatures
as compared to that in high quality bulk single crystals.
Here we have succeeded in finding and making a particular class of substrates
that allows the growth of magnetite thin films with the Verwey transition as sharp
as in the bulk. Moreover, we are now able to tune the transition temperature and,
using tensile strain, increase it to substantially higher values than in the bulk.


\end{Natabstract}

While thousands of studies have been devoted to try to understand
the first order Verwey transition in magnetite,
the high Curie temperature ($T_{C}\sim 860$~K) and the half-metallic character
of Fe$_{3}$O$_{4}$ as predicted by band theory \cite{Groot83,Yanase84} have
triggered considerable research efforts world wide to make this material suitable for
spintronic applications in the form of thin film devices
\cite{Li98apl,Greullet08,ChambersRep2000,Ziese2002,Moussy2013}. Using a variety
of deposition methods, epitaxial growth on a number of substrates has been achieved
\cite{ChambersRep2000,Ziese2002,Moussy2013,Mijiritskii2001}.
Yet, it is remarkable that in the twenty years of research on Fe$_{3}$O$_{4}$ thin films,
the first order Verwey transition \cite{Verwey39} in thin films is always broad
\cite{Ziese2002,Moussy2013,Gong97,Eerenstein02,Arora05,Geprags13}.
While in the bulk the transition takes place well within 1 K, the reported resistivity
curves for the thin films showed transition widths of about 10 K or more.
The Verwey transition temperature \textit{T}$_V$ in thin films is also much lower, with
reported values ranging from 100 to 120~K \cite{Ziese2002,Moussy2013,Gong97,Eerenstein02,Arora05,Geprags13}
while the stoichiometric bulk has \textit{T}$_V$ of 124 K.

It has been reported that several factors can affect negatively the Verwey transition in
bulk magnetite, such as oxygen off-stoichiometry \cite{Aragon93} and cation substitution
\cite{Brabers98}. The \textit{T}$_V$ gradually decreases and the transition
is claimed to change from a first order to a second or even higher order with increasing
oxygen off-stoichiometry or cation substitution. Recently, we have carried out a systematic
study on the influence of oxygen stoichiometry for the properties of magnetite thin films
\cite{Liu2014}, and we found that even for the optimal oxygen composition the
transition remains broad. In that study, we also discovered that the microstructure
of the films play an important role. In particular, with the films having a
distribution of domain sizes, a larger spread of the distribution results in a broader
transition and a small domain size gives lower transition temperatures. The transition
itself is still first order since it shows hysteresis, and there are indications that each domain has
its own transition temperature \cite{Liu2014}. Various substrates have been used in
the literature to grow epitaxial Fe$_{3}$O$_{4}$ thin films, e.g. MgO,
MgAl$_{2}$O$_{4}$, Al$_{2}$O$_{3}$, SrTiO$_{3}$, and BaTiO$_{3}$
\cite{Ziese2002,Moussy2013,Gong97,Eerenstein02,Arora05,Geprags13,Liu2014}.
These studies may suggest that the larger the lattice mismatch, the broader the transition
and the lower the average transition temperature \cite{Liu2014}.

Here we have succeeded in finding and making a particular class of substrates
that allows the growth of magnetite thin films with the Verwey transition as sharp
as in the bulk. The key principle is to obtain thin films with sufficiently large domains
and small domain size distribution. Moreover, using tensile strain we now are able to increase
the transition temperature to considerably higher values than that of the bulk.
The occurrence of the Verwey transition in the highly anisotropic
strained films raises a new question to the intricacies of the interplay between the
charge and orbital degrees of freedom of the Fe ions in magnetite,
adding another aspect of the elusiveness of this quantum material.

\vspace{5mm}
\section*{Results}

In our quest for substrates that allow for the growth of Fe$_{3}$O$_{4}$ thin films
with large domains and a narrow distribution of domain sizes, we aim first of all for
substrates with a very small lattice mismatch. Although MgO is ideal in this respect,
the occurrence of anti-phase boundaries \cite{Eerenstein02}, which
cannot be avoided when growing a (inverse) spinel film on a rocksalt substrate,
has a negative effect on the distribution and size of the domains \cite{Liu2014}.
We therefore restrict ourselves to substrates with the spinel structure. We have
identified Co$_2$TiO$_4$ with a lattice mismatch of +0.66\% as a potential candidate,
and managed to prepare large single crystals of this compound using a mirror
furnace. Substrates with $\approx$ 6 mm $\times$ 6 mm epi-polished surfaces have been made
out of these crystals. We have also prepared crystals and substrates with somewhat
larger lattice mismatch, up to +1.11 \%, by partial substitution
of the Co with Mn and/or Fe: Co$_{1.75}$Mn$_{0.25}$TiO$_{4}$ and
Co$_{1.25}$Fe$_{0.5}$Mn$_{0.25}$TiO$_{4}$. Details about the preparation and
properties of the substrate single crystals are given in the Supplementary Information
(Figures S1, S2, and S3)

Figures 1~a-d display the reflection high energy electron diffraction (RHEED) patterns of
the clean MgO~(001), Co$_2$TiO$_4$~(001), Co$_{1.75}$Mn$_{0.25}$TiO$_{4}$~(001),
and Co$_{1.25}$Fe$_{0.5}$Mn$_{0.25}$TiO$_{4}$~(001) substrates, respectively.
Figures 1~e-h show the RHEED patterns and Figures 1~i-l present the low energy electron
diffraction (LEED) of the 40 nm thick Fe$_{3}$O$_{4}$ films grown on these respective
substrates. Details of the substrate cleaning procedure as well as of the growth conditions
by using the molecular beam epitaxy (MBE) technique can be found in the Methods.
The sharpness of the RHEED stripes and the presence of Kikuchi lines, as well the
high contrast and sharpness of the LEED spots indicate a flat and well ordered single
crystalline (001) surface structure of the films. The presence of the
($\sqrt{2}\times\sqrt{2}$)R45$^{\circ}$ surface reconstruction pattern both in the RHEED
and LEED images provides another indication for the high structural quality of the films.
The valence states of the Fe ions were investigated by Fe~${2p}$ core level and valence
band x-ray photoelectron spectroscopy (XPS), thereby showing the typical signatures of
stoichiometric magnetite, see Fig. S4 in the Supplementary Information. We note that
the XPS spectra of the films grown on different substrates are identical, meaning that
the substrate has no influence on the chemical composition of the films.

Figure 2a shows the temperature dependence of the resistivity of 40 nm-thick
Fe$_{3}$O$_{4}$ films grown on MgO (001), Co$_2$TiO$_4$ (001),
Co$_{1.75}$Mn$_{0.25}$TiO$_{4}$ (001), and
Co$_{1.25}$Fe$_{0.5}$Mn$_{0.25}$TiO$_{4}$ (001), as well that of a bulk single
crystal Fe$_{3}$O$_{4}$. We can clearly see that the bulk sample has a sharp Verwey
transition at 124 K while the film on MgO shows the typical broad transition at lower
temperatures. In contrast, the thin films grown on the spinel substrates all show a
very sharp transition, almost as sharp as the bulk. The hysteresis is all within 0.7 K.
Most exciting is that now the transition temperatures of these thin films are even
higher than that of the bulk. Defining \textit{T}$_{V+}$ (\textit{T}$_{V-}$) as
the temperature of the maximum slope of the \textit{$\rho$}~(\textit{T}) curve for
the warming up (cooling down) branch, the \textit{T}$_{V+}$  is 127 K for the
film grown on Co$_2$TiO$_4$~(001) (lattice mismatch +0.66$\%$), 133 K for
the film on Co$_{1.75}$Mn$_{0.25}$TiO$_{4}$ (001) (+0.98$\%$), and even 136 K for
the film on Co$_{1.25}$Fe$_{0.5}$Mn$_{0.25}$TiO$_{4}$~(001) (+1.11$\%$), which
is 12 K higher than the \textit{T}$_{V+}$ of the bulk and about 15-35 K higher than
the \textit{T}$_{V+}$ of films of similar thickness reported in the literature so far
\cite{Ziese2002,Moussy2013,Gong97,Eerenstein02,Arora05,Geprags13,Liu2014}.

Figure 2b plots \textit{T}$_{V+}$ as a function of the film thickness on the three variants
of the spinel substrates, as well that on the rocksalt MgO.
The black horizontal line represents the \textit{T}$_{V+}$ of the
bulk single crystal. We can observe that \textit{T}$_{V+}$ gradually increases with the
film thickness, and again, that it is larger for the spinel substrates with the larger lattice
constant mismatch between the film and the substrate. While films with thicknesses
of 5 nm and less do not show a Verwey transition, we can clearly see a well defined
transition for films when they are 10 nm or thicker. In fact we would like to note that films
as thin as 10 nm grown on Co$_{1.75}$Mn$_{0.25}$TiO$_{4}$ (001) (+0.98$\%$) and
Co$_{1.25}$Fe$_{0.5}$Mn$_{0.25}$TiO$_{4}$~(001) (+1.11$\%$) have a
\textit{T}$_{V+}$ which is already comparable to that of the bulk, a highly remarkable
result in view of the generally low values even for thicker films known in the literature
\cite{Ziese2002,Moussy2013,Gong97,Eerenstein02,Arora05,Geprags13,Liu2014}.

We now investigate the strain state of the Fe$_{3}$O$_{4}$ films
grown on the spinel substrates using x-ray diffraction (XRD). Figure 3 presents the
reciprocal space mapping of the (115) reflection from a 200~nm-thick film grown on
Co$_2$TiO$_4$ (001) (+0.66$\%$) (panel a), and from a 80~nm-thick film grown on
Co$_{1.25}$Fe$_{0.5}$Mn$_{0.25}$TiO$_{4}$ (001) (+1.11$\%$) (panel b). The well
aligned longitudinally (115) reflection of the film and the substrate demonstrates that
the films are fully strained. Long range $\theta$ - 2$\theta$ XRD data indicate that there are no other phases than
Fe$_{3}$O$_{4}$. We summarize the lattice constants of three variants of the spinel
substrates, together with those of the 40 nm, 80 nm and 200 nm Fe$_{3}$O$_{4}$ films
grown on top of these substrates in table 1 from the XRD measurements. As the lattice
constant \textbf{a} of the substrates increases, i.e. 8.4528 \AA, 8.4797 \AA, and 8.4898 \AA,
the lattice constant \textbf{c} of the films gradually decreases, e.g. 8.3332 \AA,
8.3042 \AA, 8.2800 \AA, respectively, for the 40 nm films. To compare, the lattice constant of
bulk Fe$_3$O$_4$ is 8.396 \AA. From this we calculate that the Poission ratio is in the
range of 0.36-0.38, and that the volume increases as 595.29, 597.20, and
597.19~\textrm{\AA}$^{3}$, respectively, while the volume of bulk Fe$_3$O$_4$
is about 591.86~\textrm{\AA}$^{3}$.

We can also get an insight about the microstructure of the films by analyzing the peak
profile of the rocking curves of the (115) reflection. From the inverse of the
peak width, we can estimate that the average domain size is about 46 nm, 61 nm, and
68 nm for the 40 nm film grown on the three spinel substrates, respectively. These
numbers are higher than the 30 nm value found for an equivalent thick film grown on
MgO \cite{Liu2014}, suggesting that the spinel structure of the substrate indeed does help
to obtain films of structurally better quality. These results therefore support our conjecture
that one needs films with sufficiently large domains and sufficiently narrow distribution
of domain sizes in order to obtain sharp first order transitions. Perhaps this
is the solution to remedy the broad first order transitions observed in, for example,
V$_2$O$_3$ thin films
\cite{Imada1998,Brockman2012,Sakai2015},
and RENiO$_{3}$ (RE = Pr, Nd, Sm) thin films
\cite{Imada1998,Hepting2014,Mikheev2015,Zhang2016,Shi2013},
e.g. one has to carefully design substrates with a matching lattice structure
and sufficiently small lattice mismatch.

\vspace{8mm}
\section*{Discussion}

Having achieved magnetite thin films with a Verwey transition as sharp as the bulk,
we now are ready to meaningfully measure and analyze the effect of the strain exerted
by the substrate on the transition of the film. It has been reported that the Verwey transition
of bulk magnetite becomes broad and that \textit{T}$_{V}$ drops linearly with increasing
applied hydrostatic pressure and corresponding decrease of the unit cell volume
\cite{Nakagiri86,Ramasesha94,Rozenberg96,Brabers1999}.
While the application of negative hydrostatic pressures is experimentally out of reach,
our finding that the \textit{T}$_{V}$ of the Fe$_{3}$O$_{4}$ thin films increases
when epitaxially grown with increasing unit cell volume indicates that we have
in fact succeeded to exert effectively negative pressures on magnetite by using the tensile
strain imposed by the carefully chosen spinel substrates. Viewing the Verwey transition
as a transition from a Wigner crystal to a Wigner glass of small polarons \cite{Mott1967},
one can readily accept that changing the one-electron band width, and therefore
also the polaron band width, will alter the transition temperature \cite{Cullen1973}.
In particular, enlarging the lattice constant and inter-atomic distances will facilitate
the formation of an ordered state in which the different lattice sites have different
local valence and orbital states.

Yet it is important to note that the negative pressures exerted on these thin films
are by no means isotropic and therefore cannot be considered as being the equivalent
of negative hydrostatic pressures. On the contrary, the films are expanded in the plane
but compressed along the c-axis direction. This makes the Verwey transition in the
tensile strained films even more interesting: would the charge and orbital order be of
the same type as in the bulk, for which there is a lot of debate,
see Refs. \cite{Wright2001}, \cite{Senn2012}, \cite {Tanaka2013}, and references therein.
In this respect, we would like to note that the
resistivity across the Verwey transition changes by about a factor 10 in the films
while it is by a factor 100 in the bulk. Clearly the presence of the substrate limits
a full opening of the conductivity gap by putting constraints on the crystal structure.
Nevertheless the transition does take place and so the question arises what are
the minimal conditions required in terms of the local electronic structure such as
the orbital state for such a transition to occur. It would be extremely interesting
to address this question by, for example, carrying out charge and/or orbital
sensitive resonant x-ray diffraction experiments,
see Ref. \cite{Tanaka2013} and references therein.

\vspace{8mm}
\section*{Methods}

Fe$_{3}$O$_{4}$ thin films were grown by using molecular beam epitaxy (MBE) in an ultra-high vacuum (UHV) chamber with a base pressure in the high 10$^{-11}$ mbar range. High purity Fe metal was evaporated from a LUXEL Radak effusion cell at temperatures of about 1250 $^{\circ}\mathrm{C}$ in a pure oxygen atmosphere onto single crystalline Co$_2$TiO$_4$~(001), Co$_{1.75}$Mn$_{0.25}$TiO$_{4}$~(001), and Co$_{1.25}$Fe$_{0.5}$Mn$_{0.25}$TiO$_{4}$~(001) substrates. These substrates were annealed for two hours at 400 $^{\circ}\mathrm{C}$ in an oxygen pressure of 3$\times$10$^{-7}$ mbar to obtain a clean and well-ordered surface structure prior to the Fe$_{3}$O$_{4}$ deposition. The substrate temperature was kept at 250 $^{\circ}\mathrm{C}$ in an oxygen pressure of 1$\times$10$^{-6}$ mbar during growth. The Fe flux was calibrated using a quartz-crystal monitor at the growth position prior to deposition and set to 1~{\AA} per minute for the growth of all films.
\textit{In situ} and \textit{real-time} monitoring of the epitaxial growth was performed by reflection high-energy electron diffraction (RHEED) measurements, which were taken at 20~keV electron energy, with the beam aligned parallel to the [100] direction of the substrate. The crystalline structure was also verified \textit{in situ} after the growth by low-energy electron diffraction (LEED), which was recorded at the electron energy of 88~eV. All samples were analyzed \textit{in situ} by x-ray photoelectron spectroscopy (XPS). The XPS data were collected using 1486.6~eV photons (monochromatized Al $K_{\alpha}$ light) in a normal emission geometry and at room temperature using a Scienta R3000 electron energy analyzer. The overall energy resolution was set to be about 0.3 eV. X-ray diffraction (XRD) was employed for further \textit{ex situ} investigation of the structural quality and the microstructure of the films. The XRD measurements were performed with a high resolution Philips X'Pert MRD diffractometer using monochromatic Cu $K_{\alpha 1}$  radiation ($\lambda$ = 1.54056~\AA).
Temperature dependent resistivity ($\rho$) measurements of the Fe$_{3}$O$_{4}$ thin films were performed by the standard four probe technique using a Physical Property Measurement System (PPMS). The $\rho$-\textit{T} curves for all the samples were measured with a current source of 0.1 $\mu$A from 300 to 80 K and back to 300 K at zero field.

\clearpage

\section*{Acknowledgments}

The authors acknowledge valuable comments from S. Wirth.
The research of X.~H.~L. was supported by the Max Planck-POSTECH Center for Complex Phase Materials.

\section*{Competing financial interests}
The authors declare no competing financial interests.

\section*{Author contributions}

X.~H.~L. grew and characterized the films with the help from C.~F.~C. and A.~D.~R..
X.~H.~L. carried out the electrical transport and magnetic properties measurements.
A.~C.~K. grew and characterized the substrates.
X.~H.~L., C.~F.~C. and L.~H.~T. performed the data analysis and wrote the manuscript.

\clearpage

\section*{Figure legends}

\noindent {{\bf Figure 1. RHEED and LEED images.}
Representative RHEED patterns of clean {\bf a}, MgO (001), {\bf b}, Co$_2$TiO$_4$ (001), {\bf c}, Co$_{1.75}$Mn$_{0.25}$TiO$_{4}$ (001), and {\bf d}, Co$_{1.25}$Fe$_{0.5}$Mn$_{0.25}$TiO$_{4}$ (001) substrates. RHEED and LEED patterns, respectively, of 40 nm Fe$_{3}$O$_{4}$ films grown on MgO (001) {\bf e} and {\bf i}, on Co$_2$TiO$_4$ (001) {\bf f} and {\bf j}, on Co$_{1.75}$Mn$_{0.25}$TiO$_{4}$ (001) {\bf g} and {\bf k}, and on Co$_{1.25}$Fe$_{0.5}$Mn$_{0.25}$TiO$_{4}$ (001) {\bf h} and {\bf l}.}\\

\noindent {{\bf Figure 2. Electrical transport properties.}
{\bf a}, Resistivity as a function of temperature of 40 nm Fe$_{3}$O$_{4}$ thin films grown on MgO (001), Co$_2$TiO$_4$ (001), Co$_{1.75}$Mn$_{0.25}$TiO$_{4}$ (001), and Co$_{1.25}$Fe$_{0.5}$Mn$_{0.25}$TiO$_{4}$ (001) substrates and of single crystal bulk Fe$_{3}$O$_{4}$. {\bf b}, The Verwey transition temperature (\textit{T}$_{V+}$) as a function of film thickness of the films grown on MgO (001), Co$_2$TiO$_4$ (001), Co$_{1.75}$Mn$_{0.25}$TiO$_{4}$ (001), and Co$_{1.25}$Fe$_{0.5}$Mn$_{0.25}$TiO$_{4}$ (001) substrates, the black horizontal line represents the \textit{T}$_{V+}$ of a stoichiometric bulk magnetite crystal.}\\

\noindent {{\bf Figure 3. Reciprocal space maps.}
Reciprocal space mapping of the (1 1 5)-reflection of {\bf a}, a 200 nm Fe$_{3}$O$_{4}$ film grown on Co$_2$TiO$_4$~(001), and of {\bf b}, a 80 nm Fe$_{3}$O$_{4}$ film grown on Co$_{1.25}$Fe$_{0.5}$Mn$_{0.25}$TiO$_{4}$~(001).}\\

\noindent {{\bf Table1. Lattice constants.}
Lattice constants of 40 nm, 80 nm, and 200 nm Fe$_3$O$_4$ thin films on Co$_2$TiO$_4$, Co$_{1.75}$Mn$_{0.25}$TiO$_{4}$, and Co$_{1.25}$Fe$_{0.5}$Mn$_{0.25}$TiO$_{4}$.}\\

\clearpage

\begin{figure}
    \centering
    \includegraphics[width=16cm]{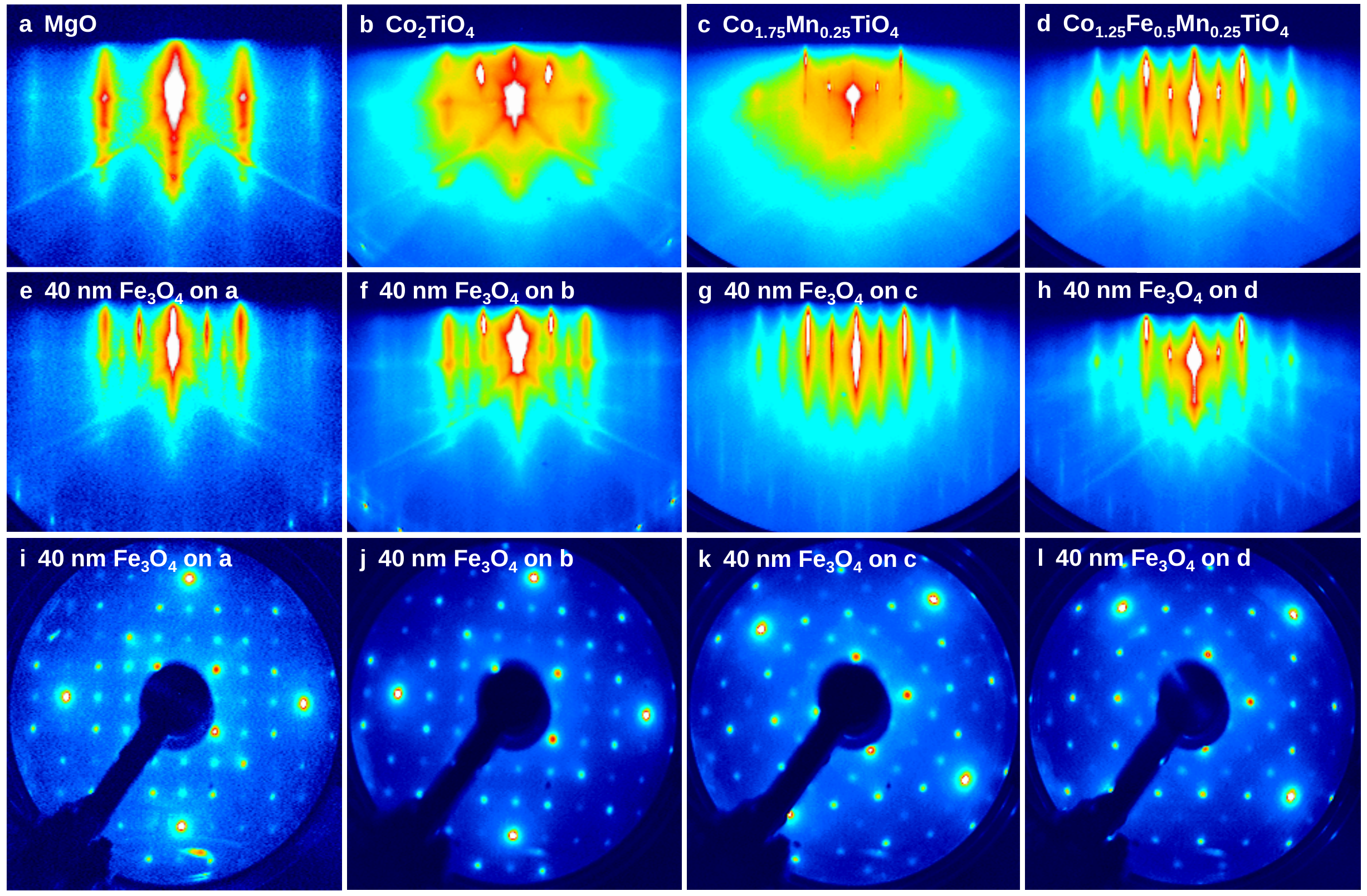}
 \caption{}
    \label{Fig1}
\end{figure}

\begin{figure}
    \centering
    \includegraphics[width=16cm]{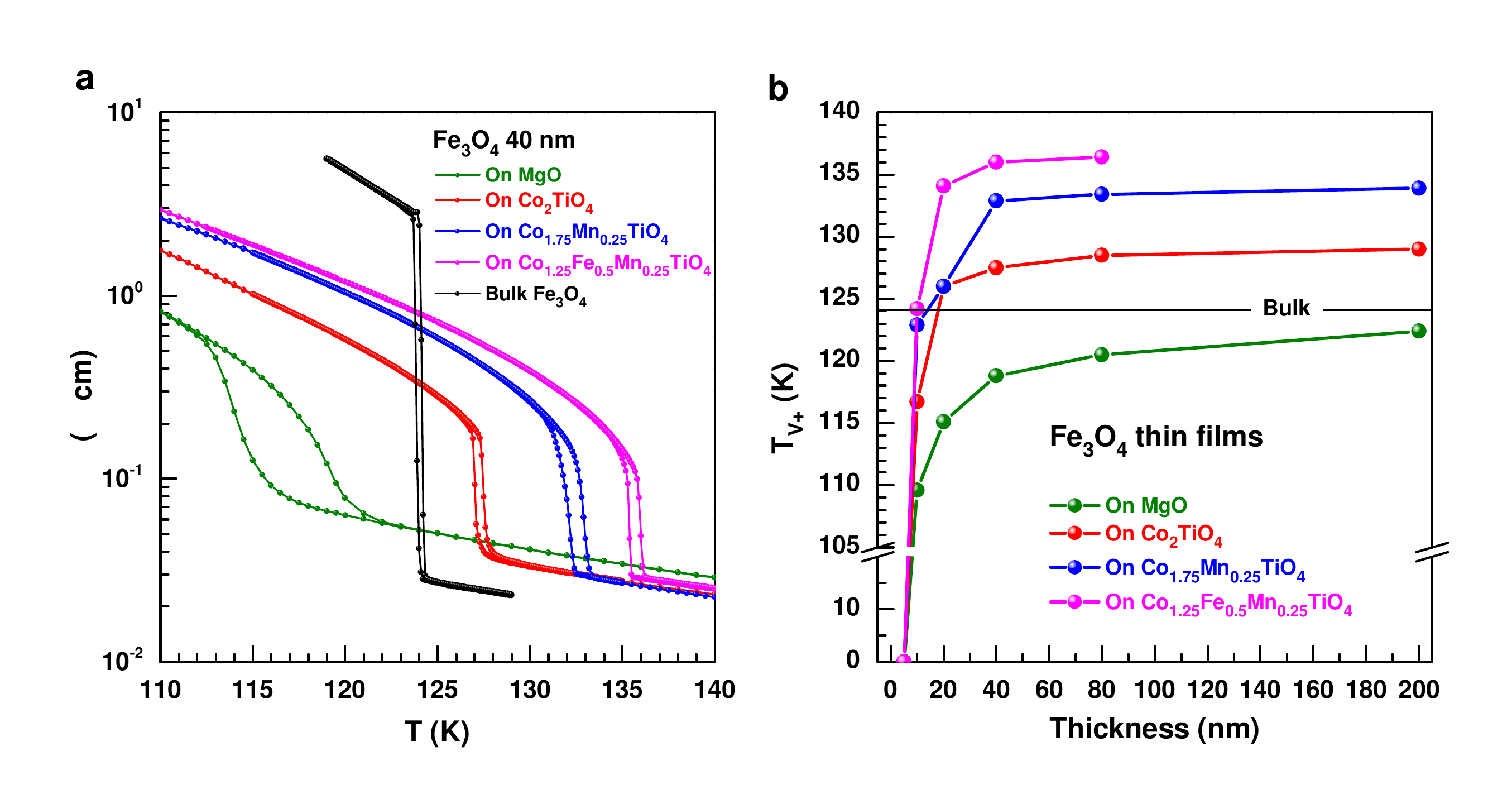}
 \caption{}
    \label{Fig2}
\end{figure}

\begin{figure}
    \centering
    \includegraphics[width=16cm]{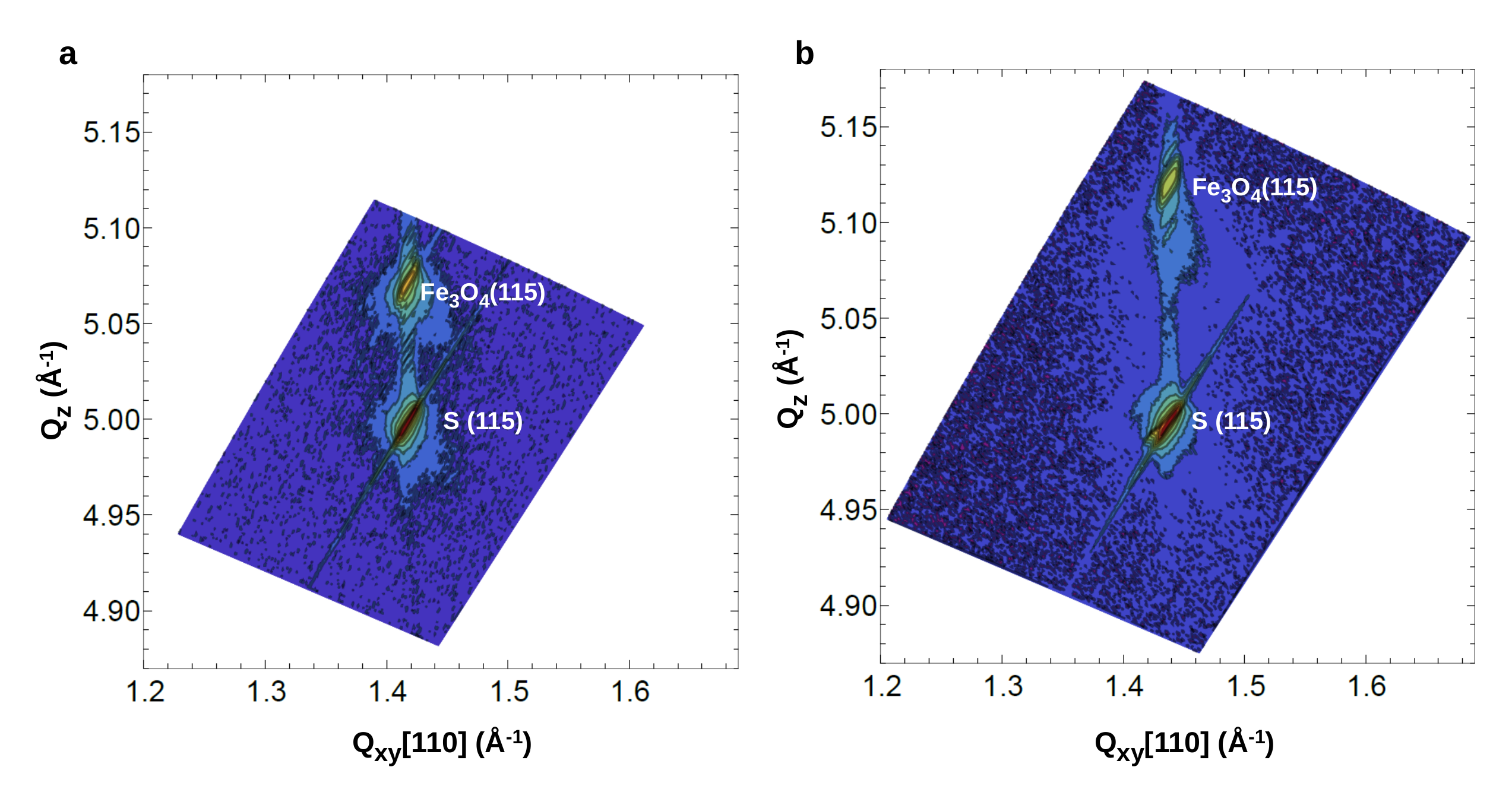}
 \caption{}
    \label{Fig3}
\end{figure}

\begin{table}
    \centering
    \includegraphics[width=16cm]{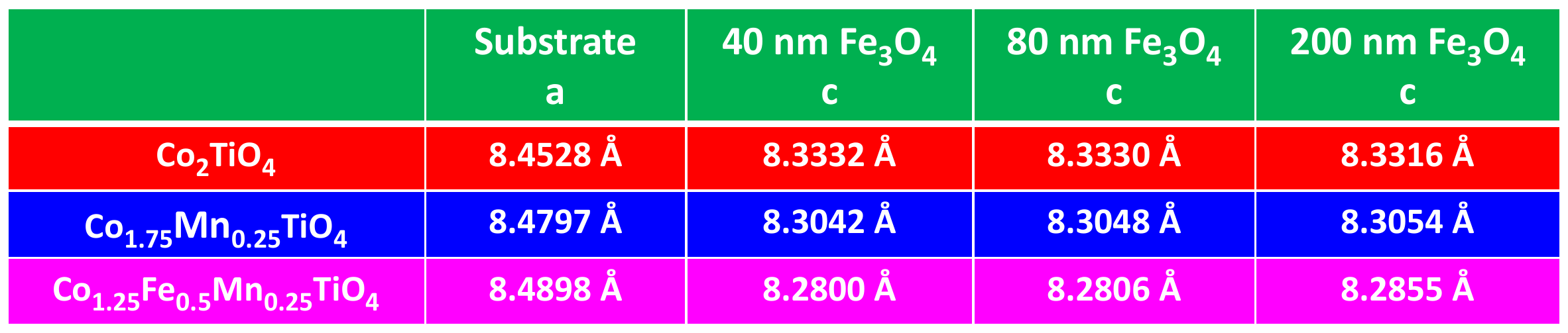}
 \caption{}
    \label{table1}
\end{table}

\clearpage
\section*{Supplementary Information}

\paragraph*{Substrates Characterization}

We confirmed the sample quality of our Co$_{2}$TiO$_{4}$ single crystals by powder X-ray diffraction, Laue diffraction and single crystal X-ray diffraction measurements as well as by complementary magnetization measurements. Co$_{2}$TiO$_{4}$ crystallizes in the cubic spinel structure AB$_{2}$O$_{4}$ with space group \textit{Fd}$\overline{3}$m. From the powder X-ray diffraction measurements (Cu-$K_{\alpha}$ radiation) it is evident that there appears to be no impurity phase within our Co$_{2}$TiO$_{4}$ crystals, see Fig. S1. The lattice parameter a of the cubic unit cell amounts to a=8.45276(20)~\AA. Laue as well as single crystal X-ray diffraction measurements confirm the single crystallinity of our crystals, see Fig. S2a. The crystal structure of a spherical Co$_{2}$TiO$_{4}$ sample has been measured by means of single crystal X-ray diffraction on a Bruker D8 VENTURE X-ray diffractometer (equipped with a bent monochromator, CMOS large area detector and a Mo-$K_{\alpha}$ X-ray tube). 4280 reflections have been measured up to 2$\Theta_{max}=135.1^{\circ}$ with a redundancy of 14.76. The internal R-value of the data collection amounts to 2.37\%. The only free structural parameter apart from the atomic displacement parameters is the x-position of the oxygen ion which amounts to x(O1)=0.26019(6). The displacement parameters amount to U$_{iso}$(Co1)=0.00641(2)~{\AA}$^2$, U$_{iso}$(Ti1/Co2)=0.006450(2)~{\AA}$^2$, U$_{iso}$(O1)=0.01131(5)~{\AA}$^2$ which are reasonably small. Finally, the refinement of the \textit{A}-site and \textit{B}-site occupancies indicates that the tetrahedral \textit{A}-site is solely occupied by the Co$^{2+}$-ions, whereas roughly a 50\%:50\% statistical distribution of Ti$^{4+}$ and Co$^{2+}$-ions is observed at the octahedral \textit{B}-sites, i.e. the refined \textit{A}-site Co$^{2+}$ occupancy amounts to 99.53(1)\%, see Fig. S2b. The goodness of the fit of this structural refinement on F$^2$ amounts to GoF=1.81, and the R- and weighted R-values amount to R=1.77\% and Rw=4.79\% respectively, thus, indicating a high reliability of our crystal structure determination.	
A complementary magnetization measurement of Co$_{2}$TiO$_{4}$ on a SQUID magnetometer (0.1 T field- and zero-field-cooled measurements) indicates a magnetic ordering temperature around 50 K with a behavior that is very similar as described in literature, \cite{HubschS} just that the ferrimagnetic ordering temperature appears to be distinctly higher (and with sharper transition) in our crystals, thus, indicating the high crystal quality of our Co$_{2}$TiO$_{4}$ single crystals, see Fig. S3.
In order to change the strain exerted by the substrate we have doped Mn and Fe into the Co$_{2}$TiO$_{4}$ matrix. In Fig. S1 the results of our Rietveld fits are shown also for these Co$_{2-x-y}$Fe$_{y}$Mn$_{x}$TiO$_{4}$ systems and indicate that there are no detectable impurity phases in our single crystals. The lattice parameters amount to a=8.45276(20)~{\AA}, 8.47974(26)~\textrm{\AA} and 8.48980(28)~\textrm{\AA} for Co$_{2}$TiO$_{4}$, Co$_{1.75}$Mn$_{0.25}$TiO$_{4}$ and Co$_{1.25}$Fe$_{0.5}$Mn$_{0.25}$TiO$_{4}$ respectively. Laue diffraction and single crystal X-ray diffraction measurements were also done on the co-doped Co$_{2-x-y}$Fe$_{y}$Mn$_{x}$TiO$_{4}$ single crystals and prove the single crystallinity of all our samples.

\clearpage

\begin{figure}
    \centering
    \includegraphics[width=12cm]{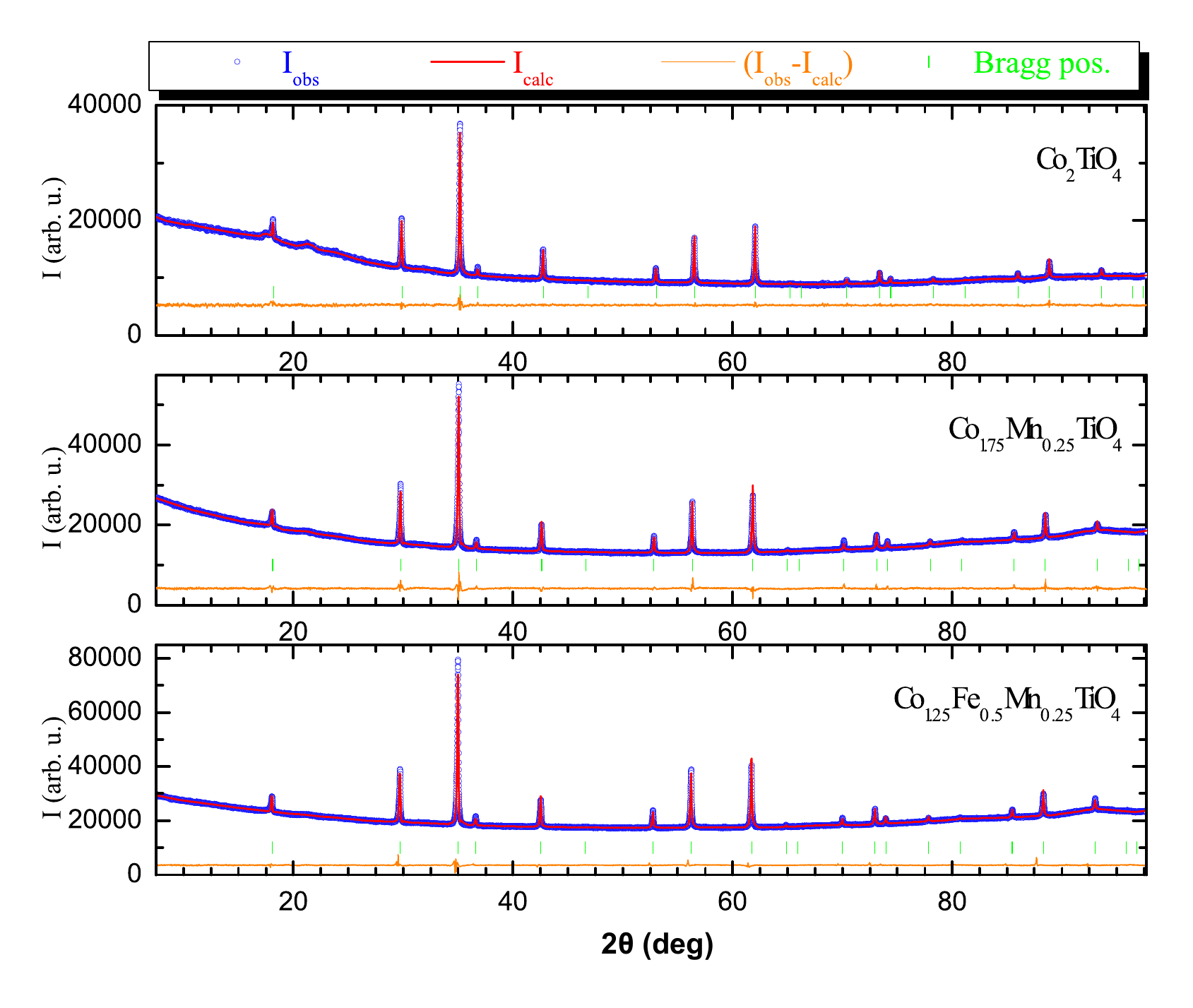}

\vspace{10pt}
\noindent {Figure S1. Rietveld refinement results of powder X-ray diffraction measurements of our Co$_{2-x-y}$Fe$_{y}$Mn$_{x}$TiO$_{4}$ crystals (obtained from single crystals crushed to powder). As indicated by the difference line (Iobs-Icalc), there are no impurity phases in our single crystals. Some minor background problems are artificial observations within our powder X-ray diffraction measurements originating from the measurement configuration and not intrinsically from our sample (around 20$^{\circ}$ - 25$^{\circ}$ in 2$\Theta$). 	
The lattice parameters amount to a= 8.45276(20) \AA, 8.47974(26) \AA, and 8.48980(28) \AA for Co$_{2}$TiO$_{4}$, Co$_{1.75}$Mn$_{0.25}$TiO$_{4}$, and Co$_{1.25}$Fe$_{0.5}$Mn$_{0.25}$TiO$_{4}$ respectively.}
\end{figure}

\begin{figure}
    \centering
    \includegraphics[width=12cm]{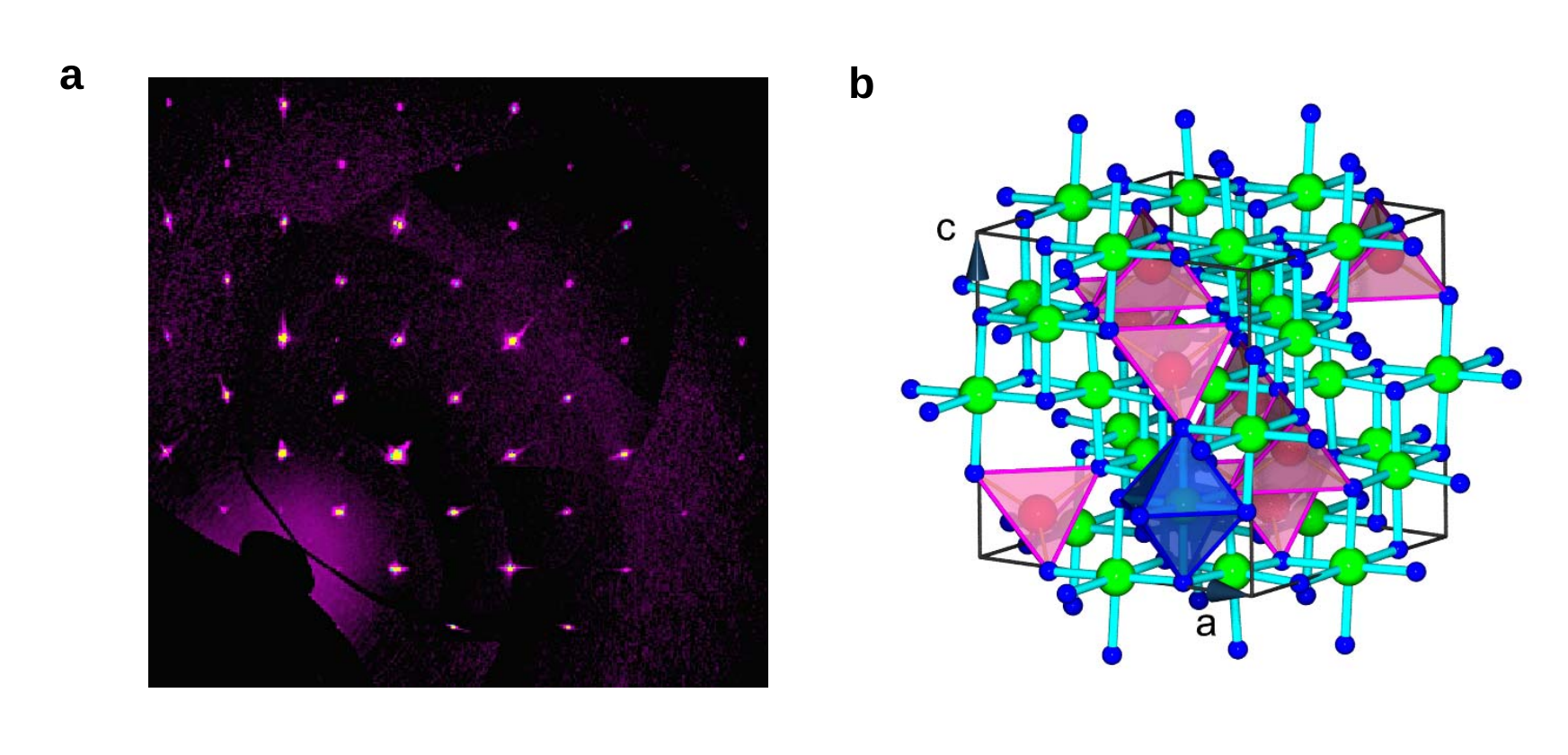}

\vspace{10pt}
\noindent {Figure S2. {\bf a}, X-ray scattering intensities of Co$_{2}$TiO$_{4}$ within the HK0 plane of reciprocal space. Obviously, our sample is a single phase single crystal without any other crystallites. {\bf b}, Visualization of the crystal structure derived from our single crystal X-ray diffraction measurements of Co$_{2}$TiO$_{4}$. Red/blue spheres: Co-/O-ions; green spheres: 50\% Co-ions and 50\% Ti-ions.}
\end{figure}

\begin{figure}
    \centering
    \includegraphics[width=12cm]{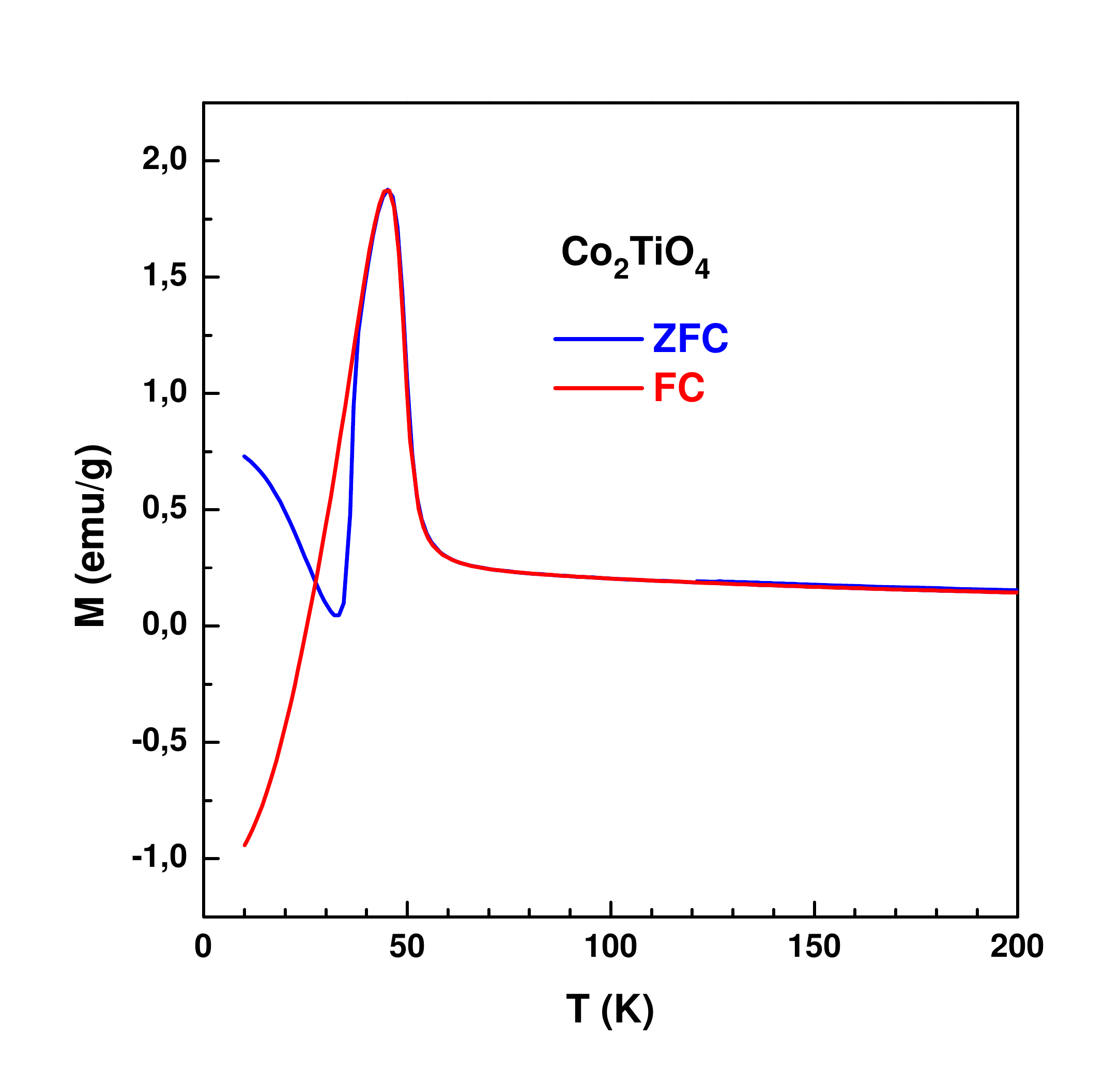}

\vspace{10pt}
\noindent {Figure S3. Magnetic susceptibility measurements of Co$_{2}$TiO$_{4}$ with an applied magnetic field of 0.1 T, confirming not only the high crystal quality indicated by the sharp magnetic transition at comparably high transition temperature, but also indicate there is no magnetic transition from the substrate in the interesting temperature regime around the Verwey transition.}
\end{figure}

\begin{figure}
    \centering
    \includegraphics[width=12cm]{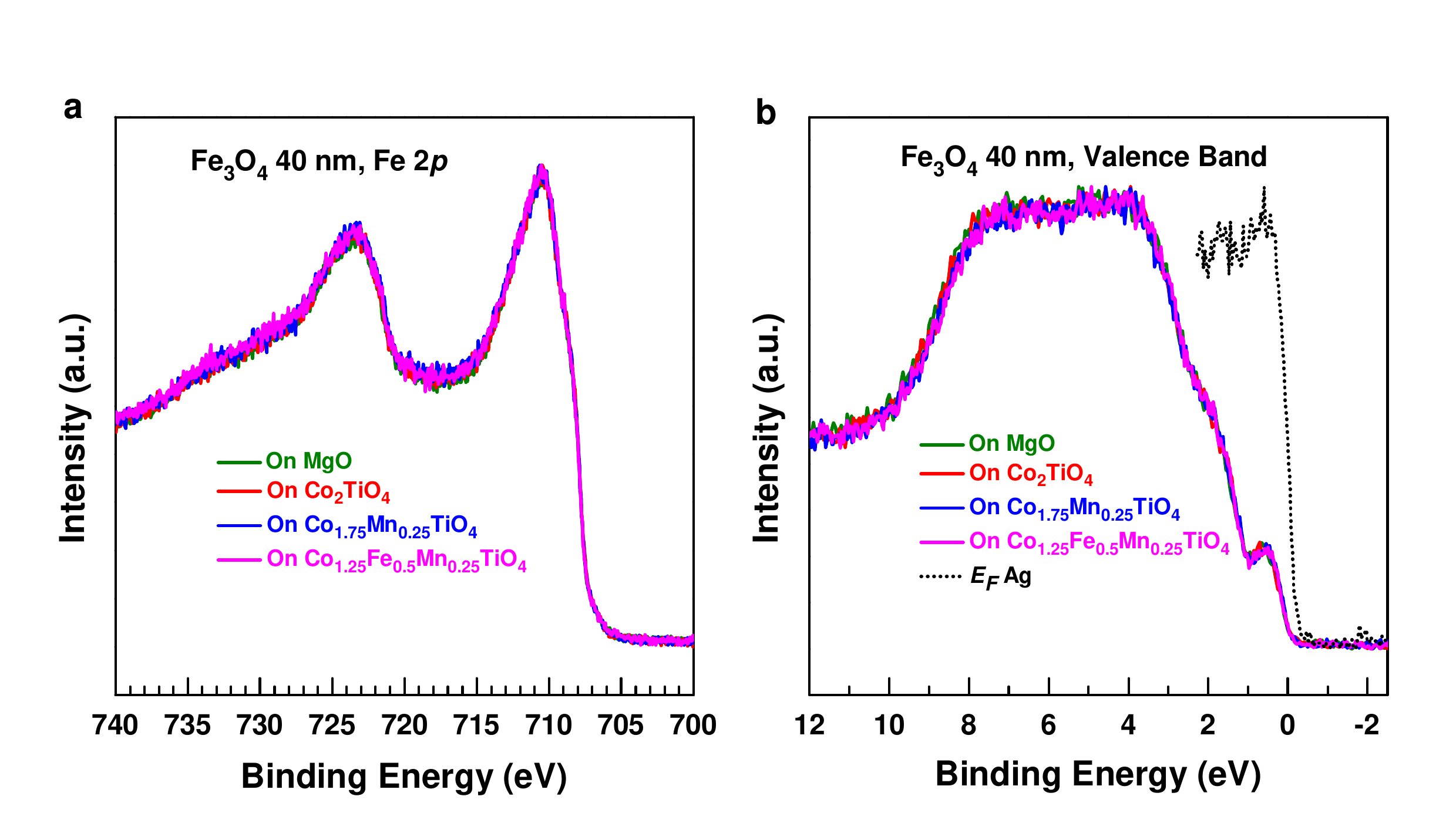}

\vspace{10pt}
\noindent {Figure S4. {\bf a}, XPS Fe~${2p}$ core-level spectra, and {\bf b}, valence band spectra of 40 nm Fe$_3$O$_4$ films on MgO, Co$_{2}$TiO$_{4}$, Co$_{1.75}$Mn$_{0.25}$TiO$_{4}$, and Co$_{1.25}$Fe$_{0.5}$Mn$_{0.25}$TiO$_{4}$.}
\end{figure}

\end{document}